\documentclass[aps,prl,showpacs,floatfix,amsmath,amssymb,
               preprint]{revtex4-1}
\usepackage{color}
\usepackage{graphicx}
\usepackage{natbib}
\usepackage{epsfig}
\usepackage{setspace}
\usepackage{amsmath}
\usepackage{amssymb}
\usepackage{verbatim}
\usepackage{bibentry}

\begin{document}
\title{Kondo Effect Route to Probing Majorana-Kramers Pairs Coupled to Quantum Dot}
\author{Rukhsan Ul Haq }
\affiliation{Quantum Data Scientist, IBM \\
Bangalore, India}
\author{ Keshav Singh}
\affiliation{Indian Institute of Science Education and Research\\
Thiruvananthapuram, India}

\begin{abstract}
In this paper, we study a set-up in which quantum dot is tunnel-coupled to Majorana-Kramers pair(MKP) which are the edge modes of time reversal invariant topological superconductor(TRITOPS),  which is an experimental realization of the time reversal invariant Kitaev chain model. Our motivation is to study the effect of Majorana-Kramers pair on Kondo fixed point and conversely the effect of Kondo interaction on Majorana-Kramers pair. This way we can capture the signatures of Majorana fermions on Kondo effect in quantum dots which can be probed feasibly in experiments. Employing the renormalization group method,  we report the first study of the signatures of Majorana-Kramers pair in the dynamical spin susceptibility of the quantum dot electrons. We not only calculate the flow equations for coupling constants but also study the renormalization of observables and especially calculate dynamical susceptibility. We propose that since these dynamical quantities have signatures of Majorana-Kramers pair, so they provide an unambiguous experimental probe for detection of Majorana fermions in TRITOPS.
\end{abstract}
\maketitle
\section{Introduction}
Majorana fermions(Majorana zero modes) have recently attracted a lot of attention in condensed matter physics community due to their special topological properties which make them a promising candidate for quantum computing. Majorana fermions make topological qubits which can store the information non-locally and hence are immune to local perturbations. Their topological protection comes from $Z_{2}$ 
parity symmetry. Majorana fermions are Ising anyons and obey non-abelian statistics. 
Though it was known that Majorana fermions exist in topological superconductors\cite{Volovik}\cite{Read} but after Kitaev\cite{Kitaev} introduced a simple Hamiltonian, it was realized that it could be experimentally realized\cite{Fu}, it gave a boost to the research on Majorana fermions. So the search for Majorana fermions in solid state systems started. Since as of now there is no conclusive evidence for the topological superconductivity in nature, it was realized that we could engineer systems which harbour Majorana fermions. The recent successes in the detection and control of the Majorana fermions\cite{Mourik}\cite{Nadj}\cite{Marcus} have created a lot of hope about   Majorana fermion based quantum computing\cite{Aasen}. However, the experimental detection of Majorana fermions is based on spectroscopic signatures, and there are physical effects which contribute to the zero-bias spectral peaks including Kondo impurities and disorder. So the search for unambiguous signatures of Majorana fermions is still on. In \cite{Golub} it was proposed that Kondo effect can be used for the detection of Majorana fermions. Kondo effect is known to have a high level of experimental tunability in quantum dots. So quantum dot systems offer a very feasible and controllable way for the detection of Majorana fermions.There are other reasons also to consider a setup where Majorana fermions are coupled to quantum dots. It was shown in\cite{Flensberg}\cite{Baranger} that Majorana fermions tunnel-coupled to conduction leads do not preserve the information in qubit and also parity time
of Majorana fermions can not be determined in those experiments. They proposed an experimental setup in which Majorana fermions are coupled to quantum dot which can be coupled to normal lead(s) as well. To explore the signatures of Majorana fermions in Kondo effect and hence the interplay      
between Majorana fermions and Kondo effect, there have been recent theoretical studies
\cite{Golub}\cite{Lee}\cite{Cheng}\cite{Chirla}. In \cite{Golub} poor man's scaling was applied to study the stability of Kondo effect. Though they find that Majorana fermions have a drastic effect on Kondo effect, they have got the same scaling equations as for the standard Kondo model. In \cite{Lee} based on numerical renormalization group(NRG)  calculations, it has been found that there is Majorana fermion induced Zeeman field 
which shifts the spectral function of one species of fermions and also
there is the contribution of Majorana fermions to differential conductance. In \cite{Chirla}  similar results have been obtained using recursive Greens function method. However, none of these studies has found a new fixed point arising due to the Majorana fermions. In \cite{Cheng} based on perturbative renormalization, slave boson mean field theory and density matrix renormalization group calculations, the authors have come to the conclusion that in the strong Majorana-Dot coupling regime, physics of the system is governed by a new fixed point rather than Kondo fixed point. The signatures of the new fixed point can be found in spin susceptibility, and they find that, quite unlike Kondo fixed point, spin susceptibility is dependent on gate-voltage(particle-hole asymmetry). Though the interplay between Majorana fermions and Kondo effect has been well studied now and essential aspects of the ensuing physics are well understood, the similar studies of  the time reveresal symmetric version of Majorana fermions called Majorana-Kramers pair are still lacking. In \cite{Lutchyn17}, based on perturbative renormalization method and slave boson mean field theory, authors have concluded that in the regime of the strong coupling to MKP, Kondo fixed point becomes unstable and low energy physics is governed by Andreev fixed point. Recetnly, there has been a lot of research interest in Majorana-Kramers pairs\cite{Zhang1}\cite{ Zhang2}\cite{Knapp}\cite{Alberto}\cite{Haima}\cite{Reeg}\cite{Haim} because the time reversal symmetric case offers more exotic ways of realizing Majorana fermion qubits. Also, the interplay between Majorana-Kramers pairs and the Kondo effect is more subtle and profound than the case of standard Majorana fermions where time reversal breaking allows the emergence of local Zeeman field which tries to suppress the Kondo spin correlations. 

In this paper, we have employed flow equation renormalization method to confirm this new fixed point. Flow equation method has recently been extensively applied to study Kondo models, both in equilibrium and in non-equilibrium \cite{Kehrein}. Flow equation method is one of the few renormalization methods which captures analytically,the full cross-over from weak coupling to strong coupling regimes of the Kondo(and Anderson impurity) model. Combined with bosonization, flow equation method gives access to the Tolouse point of the Kondo model where the flow equation can bee solved analytically to get closed form solution.  We have calculated the flow equations for the parameters of the effective Hamiltonian and solved them numerically. Ours is the first study of calculating the renormalization effects of MKP on dynamical spin susceptibility and hence we propose that these signatures of MKP in spin response should be feasibly probed in experiments on quantum dots.

The rest of this paper is organized in following way. First, we consider the Hamiltonian for the system in which a quantum dot is coupled to a normal lead on one side and is side-coupled to TRITOPS. In the long chain limit, there is only one Majorana-Kramers pair in the topological superconductor. We calculate the effective Hamiltonian, Majorana-Kondo model, which has both Kondo interaction and terms arising from the coupling to Majorana fermion. To find the stability of Kondo fixed point in the presence of Majorana fermions, we calculate the flow equations for the couplings constants of the Majorana-Kramers-Kondo model. We solve the flow equations numerically. Then we calculate the flow equations of Kondo spin observable and solving the flow equations numerically, we calculate dynamic spin susceptibility, and we find clear signatures of MKP  in this key observable quantity. Once again we explore both the cases, with particle-hole symmetry being present and absent. Finally, we summarize our conclusions and discuss the interplay between Majorana fermions and Kondo effect and hence point to the unambiguous ways for the detection of Majorana-Kramers pair.
\section{ Hamiltonian for Normal Lead-Quantum Dot-TRITOPS System}
We consider a set-up in which a quantum dot is connected to a topological superconductor on one side and to a normal lead on another side. Topological superconductor is an experimental realization of Kitaev chain model and hence has two Majorana fermions on the edges of the chain. We work in the long chain limit and couple only one of the Majorana fermions to the quantum dot. The Hamiltonian for this system is:
\begin{equation}
H=\sum_{k\sigma}\epsilon_{k}c^{\dag}_{k\sigma}c_{k\sigma}^{\phantom{\dagger}}+\sum_{\sigma}\epsilon_{d}d^{\dag}_{\sigma}d_{\sigma}
+Un_{d\uparrow}n_{d\downarrow}+
\sum_{k\sigma}t_{\sigma}(c^{\dag}_{k\sigma}d_{\sigma}+h.c.)+\sum_{\sigma}i\lambda_{\sigma}\gamma(d_{\sigma}^
{\phantom{\dagger}}+
d^{\dag}_{\sigma})
\end{equation}
The first four terms constitute the Anderson impurity model for quantum dot connected to a normal lead. The last term represents the coupling of Majorana Kramers pair(MKP) with quantum dot electron, $\lambda$ being the coupling strength. We have taken general coupling, but later on, we will see that Majorana fermion gets coupled to only one species of electrons. Due to the time reversal symmetry(TRS), $t_{\uparrow}=t_{\downarrow}=t$ and $\lambda_{\uparrow}=-\lambda_{\downarrow}=\lambda$ where $t$ and $\lambda$ are taken to be real.

 Like Anderson impurity model, our Hamiltonian has many parameter regimes but to study the interplay between Majorana fermions and Kondo effect; we need to project our Hamiltonian into its Kondo regime.  We will use the projection operator method to calculate the effective Hamiltonian in the Kondo regime of our model. 
Effective Hamiltonian in singly occupied space is given by
\begin{equation}
H_{eff}=H_{11}+H_{10}\frac{1}{E-H_{00}}H_{01}+H_{12}\frac{1}{E-H_{22}}H_{21}
\end{equation}
To get the effective Hamiltonian we need to calculate the projected Hamiltonian terms $H_{12}$ and $H_{10}$. Since, there are only two tunneling terms so only these will contribute to off-diagonal part of the effective Hamiltonian.
\begin{equation}
H_{12}=\sum_{\sigma}(tc^{\dag}_{k\sigma}+i\lambda_{\sigma}\gamma)d_{\sigma}n_{d\bar{\sigma}} \quad H_{21}=H_{12}^{\dag}
\end{equation}
\begin{equation}
H_{01}=\sum_{\sigma}(tc^{\dag}_{k\sigma}+i\lambda_{\sigma}\gamma)d_{\sigma}(1-n_{d\bar{\sigma}}) \quad H_{10}=H_{01}^{\dag}
\end{equation}
To calculate the effective Hamiltonian, we need to evaluate following two terms.
\begin{align}
&H_{12}\frac{1}{E-H_{22}}H_{21}\nonumber\\
&=\sum_{\sigma}tc^{\dag}_{k\sigma}d_{\sigma}n_{d\bar{\sigma}}\frac{1}{E-H_{22}}\sum_{k'\sigma'}td^{\dag}_{\sigma'}c_{k'\sigma'}n_{d\bar{\sigma'}}
\nonumber\\
&+\sum_{k\sigma}tc^{\dag}_{k\sigma}d_{\sigma}n_{d\bar{\sigma}}\frac{1}{E-H_{22}}\sum_{k'\sigma'}i\lambda_{\sigma'}\gamma d^{\dag}_{\sigma'}c_{k'\sigma'}n_{d\bar{\sigma'}}
\nonumber\\
&+\sum_{k\sigma}i\lambda_{\sigma}\gamma d_{\sigma}n_{d\bar{\sigma}}\frac{1}{E-H_{22}}\sum_{k'\sigma'}i t_{\sigma'} d^{\dag}_{\sigma'}c_{k'\sigma'}
n_{d\bar{\sigma'}}\nonumber \\
&+\sum_{k\sigma}i\lambda_{\sigma}\gamma d_{\sigma}n_{d\bar{\sigma}}\frac{1}{E-H_{22}}\sum_{k'\sigma'}i\lambda_{\sigma'} d^{\dag}_{\sigma'}
n_{d\bar{\sigma'}}\nonumber\\
\end{align}
\begin{align}
&H_{10}\frac{1}{E-H_{00}}H_{01}\nonumber\\
&=\sum_{k\sigma}\sum_{k'\sigma'}td^{\dag}_{\sigma}c_{k\sigma}(1-n_{d\bar{\sigma}})\frac{1}{E-H_{00}}tc^{\dag}_{k'\sigma'}d_{\sigma'}(1-n_{d\bar{\sigma'}})\nonumber\\
&+\sum_{k\sigma}\sum_{k'\sigma'}td^{\dag}_{\sigma}c_{k\sigma}(1-n_{d\bar{\sigma}})\frac{1}{E-H_{00}}i\lambda_{\sigma'}\gamma d_{\sigma'}(1-n_{d\bar{\sigma'}})\nonumber\\
&+\sum_{k\sigma}\sum_{k'\sigma'}i\lambda_{\sigma}\gamma d^{\dag}_{\sigma}c_{k\sigma}(1-n_{d\bar{\sigma}})\frac{1}{E-H_{00}}tc^{\dag}_{k'\sigma'} d_{\sigma'}(1-n_{d\bar{\sigma'}})\nonumber\\
&+\sum_{k\sigma}\sum_{k'\sigma'}i\lambda_{\sigma}\gamma d^{\dag}_{\sigma}c_{k\sigma}(1-n_{d\bar{\sigma}})\frac{1}{E-H_{00}}i\lambda_{\sigma'}\gamma d_{\sigma'}(1-n_{d\bar{\sigma'}})\nonumber\\
\end{align}
Using the Abrikosov's fermion representation of spin operators, we can write the effective Hamiltonian as(also calculated in\cite{Lee}\cite{Cheng}):

\begin{align}
&H_{eff}=\sum_{k\sigma}\epsilon_{k}c_{k\sigma}^{\dag}c_{k\sigma}+\sum_{kk'} J(k,k') S.s_{kk'} -J_{0}(k)i\gamma_{\uparrow}\gamma_{\downarrow}S^{y}+
\sum_{k}J_{1}(k)i(\gamma_{\uparrow}(c_{k\uparrow}+c^{\dag}_{k\uparrow})-\gamma_{\downarrow}(c_{k\downarrow}+c^{\dag}_{k\downarrow})) \nonumber\\
&+\sum_{k}J_{2}(k) i(\gamma_{\uparrow}(c_{k\uparrow}+c^{\dag}_{k\uparrow})
+\gamma_{\downarrow}(c_{k\downarrow}+c^{\dag}_{k\downarrow}))S^{z}+
\sum_{k}J_{3}(k)i(\gamma_{\uparrow}
(c^{\dag}_{k\downarrow}S^{+}+c_{k\downarrow}S^{-})-
\gamma_{\downarrow}(c_{k\uparrow}S^{\dag}+c^{\dag}_{k\uparrow}S^{-}))
\end{align}
where $J(k,k')=t^{2}\zeta_{1+}$, $J_{0}=\lambda^{2}\zeta_{1+}$, $J_{1}=t\lambda\zeta_{-}$, 
$J_{2}=J_{3}=t\lambda\zeta_{2+}$.
\begin{align}
\zeta_{1+}=\frac{1}{\epsilon_{d}+U-\epsilon_{k}}+\frac{1}{\epsilon_{k}-\epsilon_{d}}\nonumber\\+
\frac{1}{\epsilon_{d}+U-\epsilon_{k'}}+\frac{1}{\epsilon_{k'}-\epsilon_{d}}\\
\zeta_{2+}=\frac{1}{\epsilon_{d}+U-\epsilon_{k}} +\frac{1}{\epsilon_{k}-\epsilon_{d}}\\
\zeta_{-}= \frac{1}{\epsilon_{d}+U-\epsilon_{m}} -\frac{1}{\epsilon_{m}-\epsilon_{d}}
\end{align}
\subsection{Majorana-Kramers-Kondo Model}
The effective Hamiltonian that we have obtained has terms arising from the Quantum Dot-Majorana coupling. We call this model Majorana-Kondo model (MKKM)
to differentiate it from topological Kondo model.
In Majorana-Kramers-Kondo model there are four other terms in addition to the standard Kondo interaction term. 1. Andreev Scattering term. Since there is no direct tunnelling between lead and Majorana so this term arises due to the virtual charge fluctuations.
2. In presence of time reversal symmetry, Zeeman field like term is not present in this model, however there is a term which couples $S^{y}$ of the dot electron with $S_{m}^{y}$, which is constituted from Majorana-Kramers pair. This terms makes this model different from the Majorana-Kondo model\cite{Rukhsan}\cite{Lee}\cite{Cheng}. Then there are two more terms in which impurity spin couples both to Majorana and lead electrons. 

Unlike the Majorana-Kondo model, where the particle-hole asymmetry is a relevant perturbation, in MKKM, the corresponding term is present both at particle-hole symmetry and away from it. 
\section{Flow equations for Majorana-Kramers-Kondo Model}
In this section, we will apply the flow equation renormalization group method to Majorana-Kramers-Kondo model. We will calculate the flow equations for the parameters of the model. In the case of Majorana-Kramers-Kondo model, our interest is to explore the interplay between the Kondo effect and Majorana-Kramers pair, so we will find out what happens to Kondo divergence(scale) in the presence of MKP. We will increase the QD-TSC coupling $\lambda$ and find out whether Kondo effect survives in the strong $\lambda$ regime.
To calculate the flow equation for our model, we first need to calculate the generator. Generator for the flow equations for Majorana-Kramers Kondo  model is given below:
\begin{align}
\eta=&\frac{1}{2}\sum_{pq}(\epsilon_{p}-\epsilon_{q})(J^{\uparrow}(p,q):c^{\dag}_{p\uparrow}
c_{q\uparrow}:-J^{\downarrow}(p,q):c^{\dag}_{p\downarrow}c_{q\downarrow}:)S^{z} \nonumber \\
&+\sum_{pq}(\epsilon_{p}-\epsilon_{q})J^{\perp}(p,q)
(:c^{\dag}_{p\uparrow}c_{q\downarrow}:S^{-}-
:c^{\dag}_{q\downarrow}c_{p\uparrow}:S^{+})\nonumber\\
&+\sum_{p}\epsilon_{p}J_{1}(p)i(\gamma_{\uparrow}(c^{\dag}_{p\uparrow}
-c_{p\uparrow})-\gamma_{\downarrow}(c^{\dag}_{p\downarrow}-c_{p\downarrow})) \nonumber\\
&+\sum_{p}\epsilon_{p}J_{2}(p)i(\gamma_{\uparrow}(c^{\dag}_{p\uparrow}-c_{p\uparrow})+\gamma_{\downarrow}(c^{\dag}_{k\downarrow}-c_{k\downarrow})S^{z})\nonumber \\
&+\sum_{p}\epsilon_{p}J_{3}(p)i(\gamma_{\uparrow}(c^{\dag}_{p\downarrow}S^{\dag}-c_{p\downarrow}S^{-}) -
\gamma_{\downarrow}(c^{\dag}_{k\uparrow}S^{-}-c_{k\uparrow}S^{\dag}))
\label{generator-eq}
\end{align}
\begin{align}
\eta=\eta_{K}^{\parallel}+\eta_{K}^{\perp}+\eta_{M}^{1}+
\eta_{M}^{2}+\eta_{M}^{3}
\label{generator-split-eq}
\end{align}
In equation ~\ref{generator-split-eq} flow equation generator has been written as sum of generators corresponding to 
different terms in the interaction part of the Hamiltonian.\\
To calculate the flow equations we need to evaluate the commutators of the generators with the full Hamiltonian.
Putting all the commutators together and comparing with the original Hamiltonian we obtain the flow equations for the MKKM.\\
For spin-up Kondo coupling, the flow equation is
\begin{align}
\frac{d J^{\uparrow}(p,q)}{dl}=&-(\epsilon_{p}-\epsilon_{q})^{2}
J^{\uparrow}(p,q)+\nonumber \\
&\frac{1}{2}\sum_{s}\left(2(\epsilon_{s})-(\epsilon_{p}+\epsilon_{q}))J^{\perp}(p,s)
J^{\perp}(q,s)(1-2n(s)\right)
\end{align}
Similarly for spin down Kondo Coupling, flow equation is
\begin{align}
\frac{d J^{\downarrow}(p,q)}{dl}=&-(\epsilon_{p}-\epsilon_{q})^{2}
J^{\downarrow}(p,q)+\nonumber \\
&\frac{1}{2}\sum_{s}\left(2(\epsilon_{s})-(\epsilon_{p}+\epsilon_{q}))J^{\perp}(s,p)
J^{\perp}(s,q)(1-2n(s)\right)
\end{align}
Flow equation for transverse Kondo coupling is
\begin{align}
\frac{d J^{\perp}(p,q)}{dl}=&-(\epsilon_{p}-\epsilon_{q})^{2}
J^{\perp}(p,q)+\nonumber \\
&\frac{1}{4}\sum_{s}(1-2n_{f}(s))\Big[(2\epsilon_{s}-(\epsilon_{p}+\epsilon_{q}))J^{\perp}(s,q)J^{\uparrow}(p,s))\nonumber \\
&+(2\epsilon_{s}-(\epsilon_{p}+\epsilon_{q}))(
J^{\perp}(p,s)J^{\downarrow}(q,s))\Big]
\end{align}
The flow equation for $J_{0}$ coupling constant is:
\begin{equation}
\frac{d J_{0}}{dl}=2\sum_{p}\epsilon_{p}J_{3}(p)J_{1}(p)
\end{equation}
Flow equation for Andreev scattering term is
\begin{align}
\frac{d J_{1}(p)}{dl}=&-\epsilon_{p}J_{1}(p)+\frac{1}{4}\sum_{q}(\epsilon_{p}-\epsilon_{q})J^{\uparrow}(p,q)
J_{2}(q)+\frac{1}{2}\sum_{q}(\epsilon_{p}-\epsilon_{q})J_{3}(q)J^{\perp}(p,q)\nonumber \\
&-\frac{1}{2}\sum_{q}\epsilon_{q}J_{3}(q)J^{\perp}(p,q)-\frac{1}{4}\sum_{q}\epsilon_{q}J_{2}(q)J^{\uparrow}(p,q)
\end{align}
Flow equation for Majorana coupling $J_{2}$ is:
\begin{align}
\frac{d J_{2}(p)}{dl}=&-\epsilon_{p}J_{2}(p)+\sum_{q}(\epsilon_{p}-\epsilon_{q})J^{\uparrow}(p,q)J_{1}(q)-\sum_{q}\epsilon_{q}J_{1}(q)J^{\uparrow}(p,q)\nonumber \\
&-\sum_{q}(\epsilon_{p}-\epsilon_{q})J_{3}(q)J^{\perp}(p,q)(1-2n(q))\nonumber \\
&+\sum_{q}\epsilon_{q}J_{3}(q)J^{\perp}(p,q)(1-2n(q))
\end{align}
Flow equation for Majorana coupling $J_{3}$ is:
\begin{align}
\frac{ d J_{3}(p)}{dl}=&-\epsilon_{p}J_{3}(p)-\frac{1}{2}\sum_{q}(\epsilon_{p}-\epsilon_{q})
(1-2n(q))J^{\downarrow}(p,q)J_{3}(q)\nonumber \\
&-\sum_{q}(\epsilon_{q}-\epsilon_{p})J^{\perp}(q,p)J_{1}(q)\nonumber \\
&+\frac{1}{2}\sum_{q}(\epsilon_{q}-\epsilon_{p})J^{\perp}(q,p)J_{2}(q)(1-2n(q))\nonumber \\
& -\sum_{q}\epsilon_{q}J_{1}(q)J^{\perp}(p,q) \nonumber-\frac{1}{2}\sum_{q}\epsilon_{q}J_{2}(q)J^{\perp}(q,p)(1-2n(q))\\
& -\frac{1}{2}\sum_{q}\epsilon_{q}J_{3}(q)J^{\downarrow}(p,q)
(1-2n(q))
\end{align}
\subsection{Numerical Solution of Flow Equations}
 Flow equations need to be solved numerically because they are non-linear coupled differential equations and analytical solution can be obtained only in some special limits like infrared limit where momentum dependence of coupling constants can be dropped off. In the case of MKKM, since there is a larger number of coupling constants so the system of differential equations becomes larger and even more coupled which increases the computational complexity of the problem. Also, there are more energy scales in addition to the Kondo scale. Due to the interplay of various energy scales present in the system flow equations generally become stiff. For the numerical solution of the flow equation of Majorana-Kondo model, we have taken conduction bath to have a flat density of states. Conduction band has been discretized and typically we have kept 200 energy states and since the number of flow equations to be solved scales as $O(N^{2})$ where $N$ is the number of states kept in conduction band, the dimension of systems of flow equations is of the order $10^{4}$. The computational expense, however, scales as ${\cal{O}}N^{3}$. We have used DOPRI5 which is the fifth-order Runga-Kutta method for solving ODEs. The method has proved stable for the flow equations except that due to the stiffness of the flow equations we often had to tweak the discretization of the flow parameter $l$ grid so that the lowest energy scale gets resolved.\\

Now we will present the numerical solutions of the flow equations.
\begin{figure}[h]
\includegraphics[scale=0.5]{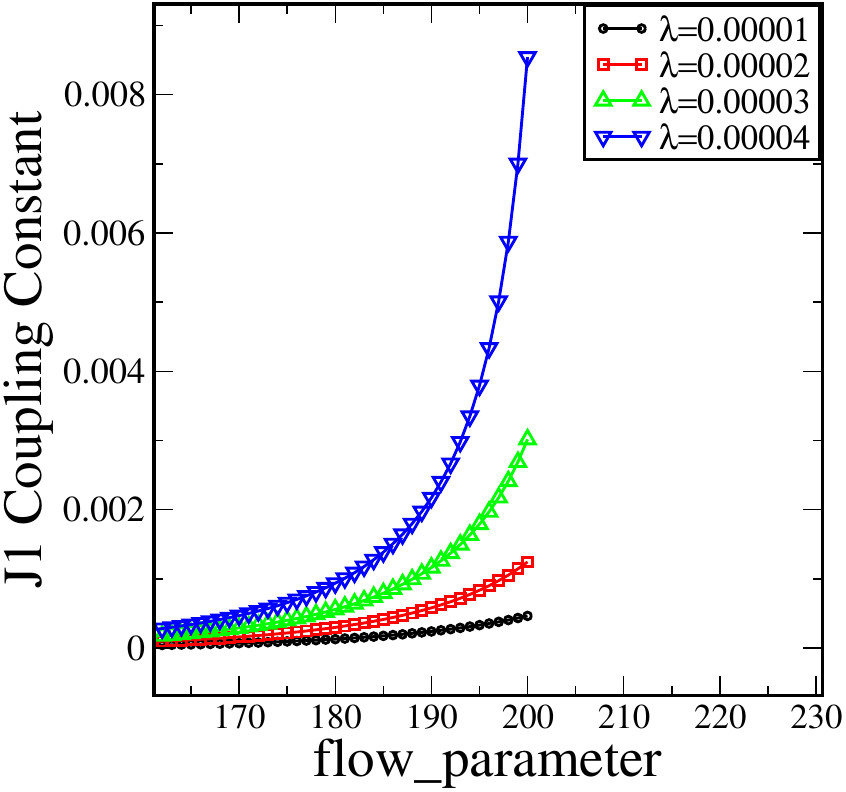}
\includegraphics[scale=0.5]{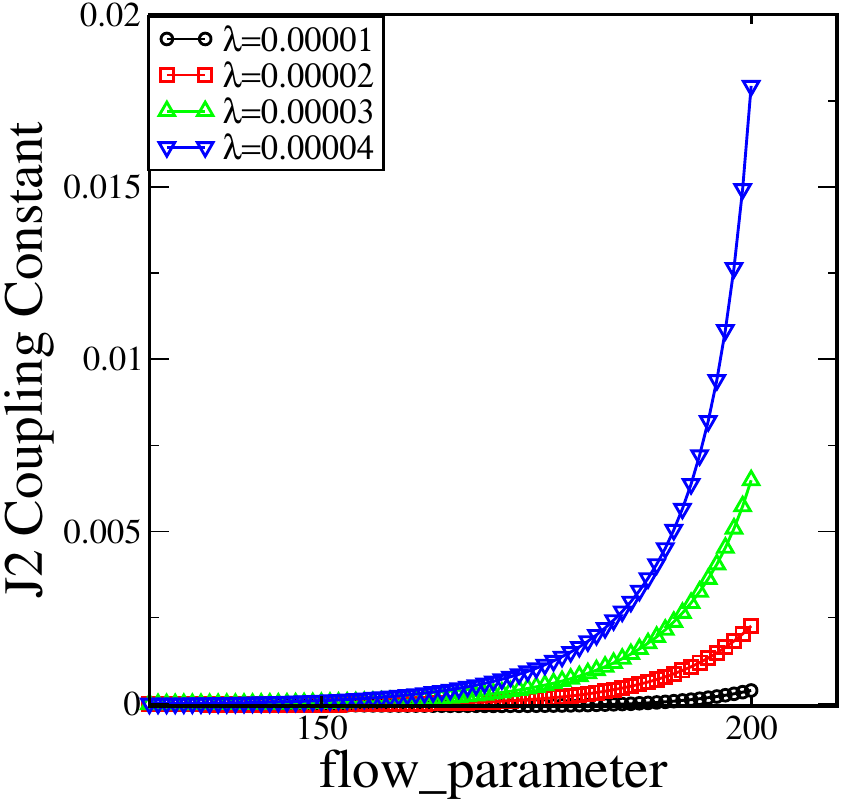}
\caption{  In left panel, $J_{1}$ coupling constant is plotted versus flow parameter. In right panel, $J_{2}$ coupling constant is plotted versus flow parameter. Other parameters are: U= 6.0, $\epsilon_{d}$= -3.0, t= 4.0.}\label{Figure 6.1}
\end{figure}
In Figure~\ref{Figure 6.1}, $J_{1}$ and $J_{2}$, the flow of the $J_{1}$ and $J_{2}$ coupling constants has been plotted. As we can see that they grow very rapidly and show the logarithmic divergence. Similar behaviour is shown by $J_{0}$ coupling constant as can been seen from the left panel of Figure~\ref{Figure 6.2}. We can clearly see that these Majorana induced coupling constants grow very fast as compared to the Kondo coupling constant, as shown in Figure~\ref{Figure 6.2}(right panel). We can conclude that in this regime of the model, there is no Kondo divergence and it is the Majorana couplings which are relevant for the low energy physics. This signifies that there may be a new fixed point which governs the low energy physics of the model as has been proposed by \cite{Lutchyn17}. It also needs to be noted that though there is no Kondo divergence, all the curves of the Kondo couplings collapse on the same curve as shown in Figure~\ref{Figure 6.1}(right panel) and hence there is not a strong renormalization of the Kondo coupling constant in this regime of the model.
\begin{figure}[h]
\includegraphics[clip=,scale=0.5]{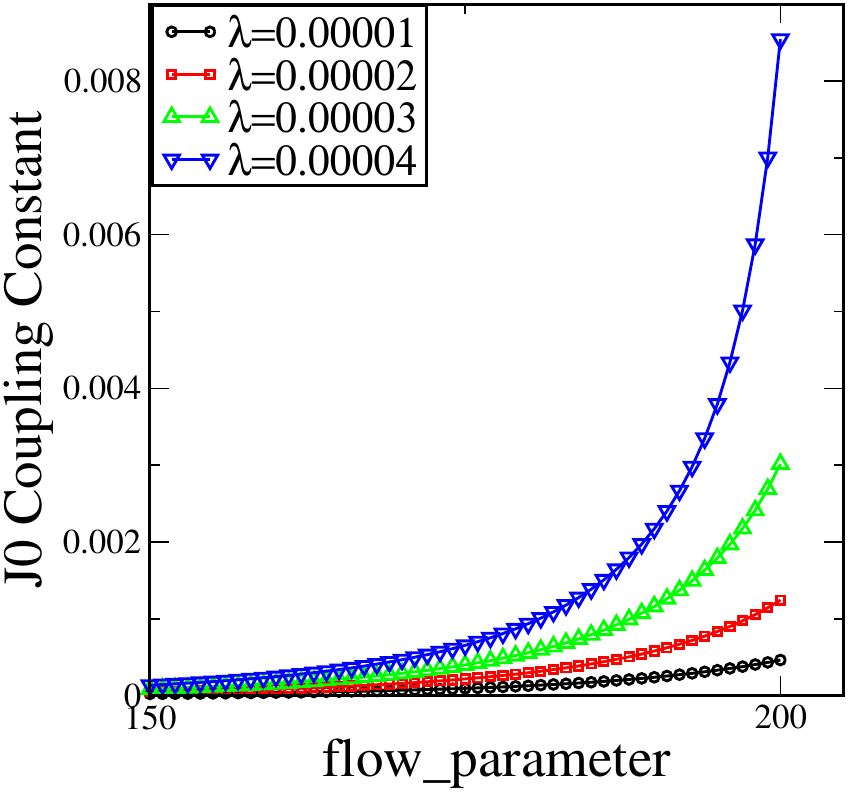}
\includegraphics[clip=,scale=0.5]{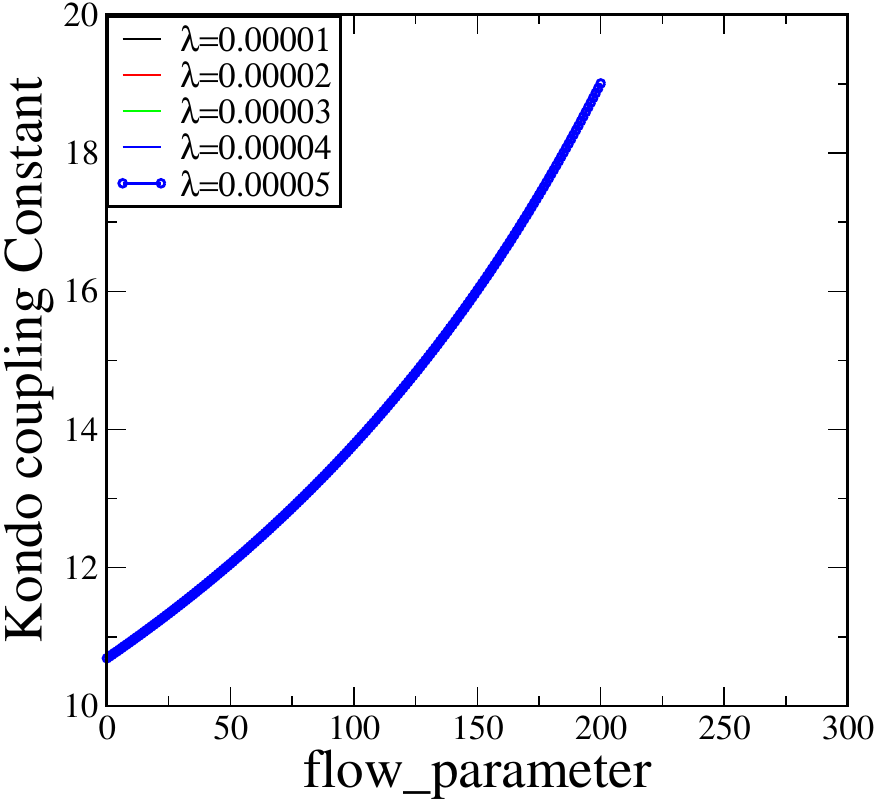}
\caption{In the left panel, flow of $J_{0}$ coupling constant has been plotted. In the right panel, flow of Kondo coupling constant has been plotted. Other parameters are same as given in ~Figure \ref{Figure 6.1} }\label{Figure 6.2}
\end{figure}


\section{Flow equations for Quantum Dot spin in Kondo regime}
One main difference between flow equations and conventional renormalization methods is that in flow equations observables also undergo unitary flow and hence renormalization. In the case of fermionic systems, fermion operators get transformed under unitary flow. The new fermion operators are then written as operator product expansion of fermion operators. In the case of the Kondo model, the spin operator gets transformed under unitary flow. To check the consistency, there are many checks which one can make using the properties of the unitary flow. Unitary flow keeps the operator algebra(commutation/anti-commutation) invariant. There are other sum rules also which can be used to check the consistency of the operator product expansion. There are physical plausibility arguments as well which can be used to make sense of these operator product expansions. In the case of the Kondo model, it is known that as renormalization flow proceeds to strong coupling fixed point, the magnetic moment of Kondo impurity gets quenched. This physical process has an algebraic mapping in the flow equation method as shown below. Using the transformed observables, we can calculate correlation and response functions for a given Hamiltonian.

 In this section, we will calculate flow equations for the Kondo impurity spin and upon solving the flow equations, obtain the dynamic spin susceptibility for the Majorana-Kramers-Kondo model. This quantity will give us access to the signatures of Majorana-Kramers pair which can be detected easily in an experiment. Hence dynamic spin susceptibility is a very important quantity to understand the interplay between MKP and Kondo effect and hence for the detection of MKP.\\
We make following ansatz for the spin operator $S^{z}$:
\begin{align}
S^{z}(l)=&h^{z}(l)S^{z}+\sum_{pq}\gamma_{pq}(l)(:c^{\dag}_{p\uparrow}
c_{q\downarrow}:S^{-}+:c^{\dag}_{q\downarrow}c_{p\uparrow}:S^{+})\nonumber \\
&+\sum_{p}\zeta_{p}(l)i(\gamma_{\uparrow}(c_{k\uparrow}+c^{\dag}_{k\uparrow})+\gamma_{\downarrow}(c_{k\downarrow}+c^{\dag}_{k\downarrow})S^{z}))\nonumber \\
&+\sum_{p}\eta_{p}(l)i(\gamma_{\uparrow}
(c^{\dag}_{k\downarrow}S^{+}+c_{k\downarrow}S^{-})-
\gamma_{\downarrow}(c_{k\uparrow}S^{\dag}+c^{\dag}_{k\uparrow}S^{-}))
\label{ansatz}
\end{align}
where $h^{z}(l=0)=1$ and initial values of all other parameters are zero. For the Kondo model, the ansatz for the spin operator\cite{Kehrein} is motivated by the fact that the Kondo effect quenches the magnetic moment. So for the isotropic Kondo model, the decomposition of the spin operator has only two terms referring to the Kondo spin and the Kondo interaction.  For the anisotropic case, since there is a Zeeman field and hence magnetization. Thus, the ansatz for spin observable needs to include these terms as well\cite{Fritsch}. For the case of Majorana-Kramers-Kondo model, due to the Majorana fermion induced couplings the Kondo impurity spin also couples to Majorana fermion. Hence these interactions need to be included in the ansatz. Andreev scattering term does not couple to the Kondo impurity spin and hence has not been included in the ansatz.\\
To calculate the flow equation of the spin operator , we will use one-loop  generator as given in equation~\ref{generator-eq}. .
\begin{align}
\frac{d S^{z}}{dl}=\left[ \eta(l),S^{z}(l)\right]
\end{align}
 In the following we will present the flow equations for the various parameters in Equation~\ref{ansatz}.\\
Flow equation for $h^{z}$ co-efficient is:
\begin{align}
\frac{d h^{z}}{dl}=&-\sum_{pq}(\epsilon_{p}-\epsilon_{q})J_{\perp}(p,q)\gamma_{pq}
(n(p)+n(q)-2n(p)n(q))
\end{align}
Flow equation for $\gamma$ co-efficient is:
\begin{align}
\frac{d \gamma_{pq}}{dl}=&\frac{1}{2}\sum_{r}\left(J^{\uparrow}(p,r)(\epsilon_{r}-\epsilon_{p})\gamma_{rq}(l)+J^{\downarrow}(r,q)(\epsilon_{r}-\epsilon_{q})\gamma_{pr}(l)\right)(1-2n(q))\nonumber\\
&+\frac{1}{2}(\epsilon_{p}-\epsilon_{q})J^{\perp}(p,q)h^{z}(l)
\end{align}
Flow equation for $\zeta$ co-efficient is:
\begin{align}
\frac{d \zeta_{p}}{dl}=&\sum_{q} \epsilon_{q}J_{3}(q)\gamma_{pq}(1-2n(q)) \nonumber\\
&-\sum_{q}(\epsilon_{p}-\epsilon_{q})J_{\perp}(p,q)\eta_{q}(1-2n(q))
\end{align}
Flow equation for $\eta$ co-efficient is:
\begin{align}
\frac{d \eta_{p}(l)}{dl}=&-\epsilon_{p}J_{3}(p)h^{z}(l)-\sum_{q}\epsilon_{q}J_{1}(q)\gamma_{qp}(l) \nonumber\\
&-\frac{1}{2}\sum_{q}\epsilon_{q}J_{2}(q)\gamma_{qp}(l)(1-2n(q))\nonumber \\
&+\frac{1}{2}\sum_{q}(\epsilon_{p}-\epsilon_{q})J_{\perp}(p,q)\zeta_{q}(l)(1-2n(q))\nonumber\\
&-\frac{1}{2}\sum_{q}(\epsilon_{p}-\epsilon_{q})J^{\downarrow}(p,q)\eta_{q}(1-2n(q))
\end{align}
\subsection{Numerical solution}
In this section we will solve the flow equations for the spin operator and calculate the dynamic spin susceptibility.   For the Majorana-Kramers-Kondo model, spin-spin correlation function is given by
\begin{align}
C(\omega)=&\frac{\pi}{4}\sum_{p}\Big(\tilde{\gamma}^{2}_{\epsilon_{p},\epsilon_{p}+\omega} n_{f}(\epsilon_{p})(1-n_{f}(\epsilon_{p}+\omega))+ n_{f}(\epsilon_{p}+\omega)(
1-n_{f}(\epsilon_{p}))\Big)
\end{align}
This correlation is related to dynamical spin susceptibility via fluctuation-dissipation theorm\cite{Kehrein}. The tilde sign denotes that $\gamma$ co-efficient is in the limit $l\rightarrow \infty$.
\begin{equation}
\chi(\omega)=tanh(\frac{\omega}{2T})C(\omega)
\end{equation}
To compute dynamic susceptibility we need to solve the flow equations for the spin observable numerically.
\begin{figure}[h]
\centering
\includegraphics[scale=0.45]{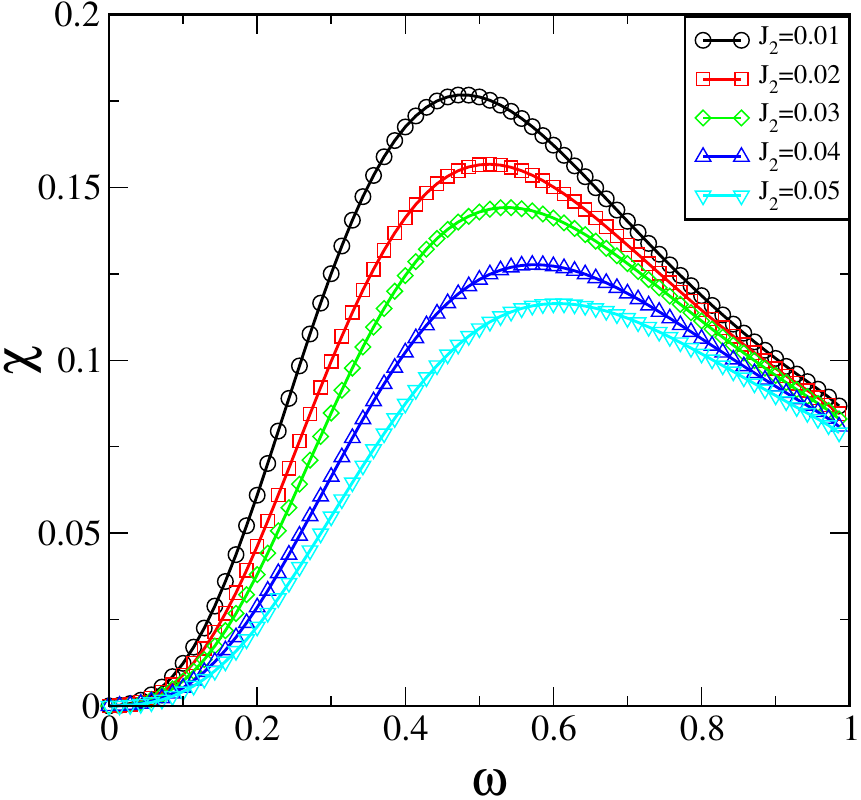}
\includegraphics[scale=0.45]{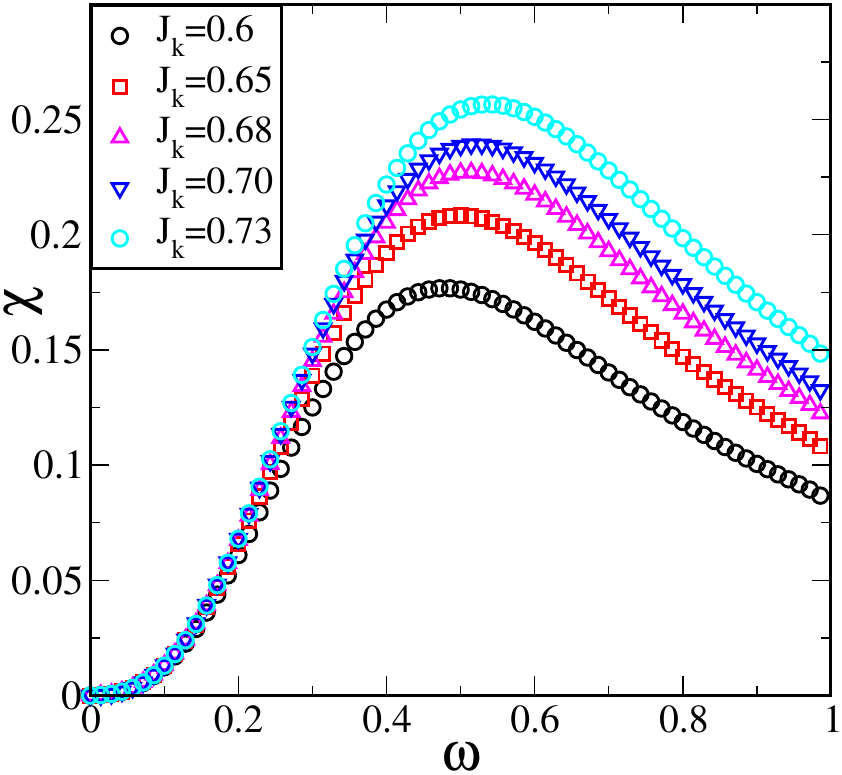}
\caption{Dynamical spin susceptibility plots for Majorana-Kramers-Kondo model : In left panel, spin susceptibility has been plotted  on frequency axis for different $J_{2}$ coupling constants. In right panel, spin susceptibility has been plotted for increasing values of Kondo coupling constant.}\label{Figure 6.5}
\end{figure}
To capture the effects of Majorana-Kramers pair in spin susceptibility, we have plotted dynamic spin susceptibility on the frequency axis in Figure~\ref{Figure 6.5}. In the left panel, we see the effect of increasing $J_{2}$ Majorana coupling. As we increase $J_{2}$, spin susceptibility decreases because of the competition from MKP induced couplings which coupled spin to the parity of the MKP. In the presence of time-reversal symmetry, there is an additional pairing which couples $S^{y}$ of the dot electron spin to the $S^{y}_{m}$ constituted by Majorana-Kramers pair. Due to its transverse nature, it gives competition to the Kondo effect and hence leads to a decrease in spin susceptibility. 

In the right panel of Figure~\ref{Figure 6.5}, we see the effect of increasing Kondo coupling constant on the spin susceptibility. We can see that the spin susceptibility gets enhanced as we increase Kondo coupling constant. This should be contrasted with the effect seen on the left panel of Figure~\ref{Figure 6.5} where we find that spin susceptibility decreases with increasing values of $J_{2}$.

\section{Conclusion}

In this paper, we have done a detailed study of the interplay between Majorana-Kramers pair and Kondo effect considering an experimental set-up in which a quantum dot is coupled to a normal lead on one side and tunnel-coupled to a TRITOPS, which itself is an experimental realization of time-reversal invariant Kitaev p-wave chain model. Our motivation comes from the recent surge in the research on Majorana fermions( and MKPs) as being the promising candidates for topological quantum computing. Hence searching for unambiguous ways of their detection has assumed utmost importance. Kondo effect in quantum dots offers a lot of tunability and hence very feasible way to detect Majorana fermions. However, to single out the signatures of Majorana fermions in Kondo physics needs careful and detailed study. There are questions which need to be addressed to have conclusive evidence for the signatures of Majorana fermions. Firstly Kondo effect is associated with the strong coupling fixed point of Anderson impurity model and is very stable to perturbations and hence to show that Majorana fermions make Kondo fixed point unstable and lead to a new fixed point, one has to take recourse to methods which can be relied upon in the strong coupling regime of the model. In our work, we have employed the flow equation method which has already been applied to the Kondo model to study its strong coupling physics both in equilibrium and non-equilibrium. Based on this method we did the renormalization group study of the Majorana-Kramers-Kondo model which is the effective Hamiltonian in the strong coupling regime of the Hamiltonian for NL-QD-TRITOPS system. We calculated the flow equations for the coupling constants of the MKKM, and after solving them numerically, we found how MKP induced terms, cut off Kondo divergence. We find that MKP induced coupling constants grow faster as compared to the Kondo coupling constant and hence they are more relevant for the low energy physics. Our calculations suggest that there may be a new fixed point determined by MKP induced coupling constants.

Dynamic spin susceptibility is an experimentally accessible quantity. We find clear signatures of MKP in this quantity and we propose these signatures can be feasibly found in experiments on quantum dots. Our studies have confirmed and consolidated the interplay between MKP and Kondo effect and hence showed that  Kondo effect in quantum dots provides an experimentally feasible and unambiguous way for the detection of Majorana-Kramers pair.
\section{Acknowledgements}
Rukhsan Ul Haq, would like to acknowledge Prof. N. S. Vidhyadhiraja for various discussions related to this work. He would also like to acknowledge Department of Science and Technology(DST GOI) for funding and JNCASR Bangalore for  the facilities and vibrant research environment.  Keshav would like to thank Department of Science
and Technology(DST GOI) for the Inspire Fellowship.
 

\begin{thebibliography}{9}
\bibitem{Volovik} G. Volovik, JETP Lett. \textbf{70}, 609(1999).
\bibitem{Read} N. Read and D. Green, Phys. Rev. B \textbf{61}, 10267 (2000).
\bibitem{Kitaev} A. Kitaev, Phys. Usp. \textbf{44}, 131 (2001).
\bibitem{Fu} L. Fu and C.L. Kane, Phys. Rev. Lett. \textbf{100}, 096407 (2008).
\bibitem{Mourik} V. Mourik \textit{et al} Science \textbf{336}, 1003 (2012).
\bibitem{Nadj} S.Nadj-Perge\textit{et al} Phys. Rev. B \textbf{88}, 020407 (2013).
\bibitem{Marcus} M.T. Deng \textit{et al} Science, Vol. 354, Issue 6319, pp.1557-1562 (2016).
\bibitem{Aasen} David Aasen\textit{et al} Phys. Rev. X \textbf{6}, 031016 (2016)
\bibitem{Flensberg} Martin Leijnse and Karsten Flensberg, Phys. Rev. B \textbf{84}, 140501 (2011).
\bibitem{Baranger} D.E. Liu and H. U. Baranger, Phys. Rev. B \textbf{84}, 201308 (2011)
\bibitem{Golub} A. Golub, I. Kuzmenko and Y. Avishai, Phys. Rev. Lett. \textbf{107}
176802 (2011)
\bibitem{Lee} M. Lee, J. S. Lim and R. Lopez, Phys. Rev. B \textbf{87}, 241402 (2013).
\bibitem{Cheng} M. Cheng, M. Becker, B. Bauer and R. M. Lutchyn, Phys. Rev. X \textbf{4}, 031051 (2014).
\bibitem{Chirla} R. Chirla, I.V. Dinu, V. Moldoveanu and C. P. Moca, Phys. Rev. B \textbf{90}, 195108 (2014).
\bibitem{Lutchyn17} Y. Kim \textit{et al}, Phys. Rev. B. \textbf{94},075439 (2016).
\bibitem{Zhang1} Rui-Xing Zhang and S. Das Sarma, arxiv:2012.13411 (2020)
\bibitem{Zhang2} Rui-Xing Zhang, William S. Cole, and S. Das Sarma, Phys. Rev. Lett. \textbf{122}, 187001 (2019)
\bibitem{Knapp} Christina Knapp, Aaron Chew,
and Jason Alicea, arXiv:2006.10772v2 (2020)
\bibitem{Alberto} Alberto Camjayi, Liliana Arrachea, Armando Aligia and Felix von Oppen, arXiv:1612.07410v2 (2017)
\bibitem{Haima} Arbel Haima and Yuval Oreg, arXiv:1809.06863v1 (2018)
\bibitem{Reeg} Christopher Reeg, Constantin Schrade, Jelena Klinovaja and Daniel Loss, Phys. Rev. B \textbf{96}, 161407(R) (2017)
\bibitem{Haim} A. Haim and Y. Oreg, Physics Reports 825, 1 (2019)
\bibitem{Kehrein} S. Kehrein \textit{The flow equation approach to many particle systems} Springer, Berlin (2006).
\bibitem{Fritsch} P. Fritsch and S. Kehrein, Phys. Rev. B. \textbf{81}, 035113 (2010).
\bibitem{Hofstetter} W. Hofstetter and S. Kehrein, Phys. Rev. B, \textbf{63}, 140402 (2001).


\end{thebibliography}
\end{document}